\documentclass[prc,twocolumn,showpacs,superscriptaddress,nofootinbib]{revtex4}
\usepackage{graphicx}
\def\be{\begin{eqnarray}}
\def\ee{\end{eqnarray}}
\def\bc{\begin{center}}
\def\ec{\end{center}}

\def\rmF{{\rm F}}
\def\rmd{{\rm d}}
\def\om{\omega}

\newcommand{\lsim}{\stackrel{\scriptstyle <}{\phantom{}_{\sim}}}
\newcommand{\gsim}{\stackrel{\scriptstyle >}{\phantom{}_{\sim}}}
\begin{document}
\title{Making a soft relativistic mean-field equation of state stiffer at high density}
\author{K. A. Maslov}
\affiliation{National Research Nuclear University (MEPhI), Kashirskoe avenue 31, RU-115409 Moscow, Russia}
\author{E. E. Kolomeitsev}
\affiliation{Matej Bel  University, Faculty of Natural Sciences, Tajovskeho 40, SK-97401 Banska Bystrica, Slovakia}
\author{D. N. Voskresensky}
\affiliation{National Research Nuclear University (MEPhI), Kashirskoe avenue 31, RU-115409 Moscow, Russia}
\begin{abstract}
We study relativistic mean-field (RMF) models including nucleons interacting with scalar, vector and iso-vector mean fields and mean-field self- and cross-interaction terms. Usually, in such models the magnitude of the scalar field increases monotonically with the nucleon density, and the nucleon effective mass decreases. We demonstrate that the latter quantity stops decreasing and  the equation of state stiffens, provided the mean-field self-interaction potential rises sharply in a narrow vicinity of the values of mean fields corresponding to  nucleon densities $n\gsim n_{*}>n_0$, where $n_0$ is the nuclear saturation density. As a result the limiting neutron star mass increases. This procedure offers a simple way to stiffen the equation of state at  densities above $n_{*}$ without altering it at densities $n\lsim n_{0}$. The developed scheme allows a neutron star application of the RMF models, which are well fitted to finite nuclei but do not fulfill the experimental constraint on the limiting neutron star mass.
The exemplary application of the method to the well-known FSUGold model allows us to increase the limiting neutron star mass from $1.72~M_\odot$ to $M \geq 2.01~M_\odot$.
\end{abstract}
\maketitle

A relativistic mean-field (RMF) model proposed and advertised in~\cite{Duerr,Walecka,SerotWalecka}
is a convenient vehicle for construction of the equation of state (EoS) of baryon matter, which preserves causality.
Various RMF models are successfully used to describe neutron stars (NS) and heavy-ion collisions, see~\cite{Glendenning,KTV,Klahn:2006ir}. The models prove to be also applicable to atomic nuclei, where a high precision in the description of gross properties of finite nuclei close to the valley of stability can be achieved~\cite{PGReinhard,Sugahara-Toki}.
An accurately calibrated  parametrization of the RMF model was introduced in~\cite{Toddrutel}, known as the FSUgold model, which allows one to compute the ground state properties of finite nuclei and neutron rich matter.

In order to describe finite nuclei well, one includes the self- and cross-interaction terms of  meson fields~\cite{Toddrutel,Mueller-Serot-96,Fattoyev10} and/or nonlinear derivative couplings~\cite{Chen14}. However these models yield a rather soft EoS and cannot describe the heavy NSs with masses of $(2.01\pm 0.04) M_\odot$, where $M_\odot$ is the mass of the Sun,
which were recently unambiguously identified experimentally~\cite{Demorest:2010bx,Antoniadis:2013pzd}.
Additional complication arises if hyperons are included in the consideration, since in their presence the EoS softens even more, cf.~\cite{Bednarek-hyp}.
To reconcile the appropriate description of the properties of atomic nuclei and the NS mass constraints, one exploits density-dependent coupling constants~\cite{Typel}. Similarly, one could use mean-field-dependent coupling constants, cf.~\cite{Kolomeitsev:2004ff,MKVshort}.

In this Rapid Communication, we  will demonstrate that, if the mean-field self-interaction potential rises sharply in a narrow vicinity of the values of mean fields corresponding to  nucleon densities $n\gsim n_{*}>n_0$, where $n_0$ is the nuclear saturation density, the nucleon effective mass saturates and  the EoS  stiffens. As a result the limiting NS mass may increase above the  value $(2.01\pm 0.04) M_\odot$.  This procedure offers a simple way to stiffen the EoS at high densities without altering it at densities $n\lsim n_{0}$.

For illustration purposes we work with the standard non-linear Walecka (NLW) model~\cite{SerotWalecka} defined by the Lagrangian
\begin{eqnarray}
\mathcal{L}&=&\overline{\Psi}_N\,
\Big[(i\,\partial_\mu - g_{\om }\om_\mu - g_{\rho }\vec{t}\vec{\rho}_\mu)\,\gamma^\mu - m_N+g_{\sigma }\sigma\Big]\,\Psi_N
\nonumber\\
&+&\frac12[\partial_\mu\sigma \partial^\mu\sigma-m_\sigma^2\sigma^2]
-\frac14\om_{\mu\nu}\om^{\mu\nu} +\frac12 m_\om^2\om_\mu\om^\mu
\nonumber\\
&-&\frac14\vec{\rho}_{\mu\nu}\vec{\rho}\,^{\mu\nu} +\frac12 m_\rho^2\vec{\rho}_\mu \vec{\rho}\,^\mu+ \mathcal{L}_{\rm int}\,,
\nonumber\\
&&\omega_{\mu\nu}=\partial_\mu \om_\nu -\partial_\nu\om_\mu\,,
\quad
\vec{\rho}_{\mu\nu}=\partial_\mu\vec{\rho}_\nu -\partial_\nu\vec{\rho}_\mu \,.
\label{Lag}
\end{eqnarray}
Here $\Psi_N=(\psi_{p},\psi_n)^{\rm T}$ is the isospin doublet of Dirac bispinors for protons ($p$) and neutrons ($n$) with the bare mass $m_N=938$~MeV; $\sigma$, $\om$, and $\vec{\rho}$ denote the fields of scalar, vector, and isovector-vector mesons with masses $m_\sigma$, $m_\om$, and $m_\rho$, respectively; $\gamma^\mu$ ($\mu=0,1,2,3$) are the standard Dirac $\gamma$ matrices;
and $\vec{t}$ stands for the operator of the baryon isospin. The $\mathcal{L}_{\rm int}$ may include the scalar and vector-field self-interaction terms and the field cross-interaction terms.   Following~\cite{Boguta77} we will first use the simplest form of the  $\mathcal{L}_{\rm int}$ reducing to the scalar-field self-interaction potential
$\mathcal{L}_{\rm int}=-U(\sigma)=-b m_N\,(g_\sigma\sigma)^3/3-c\,(g_\sigma\sigma)^4/4$; $g_{\sigma }$, $g_{\om }$, $g_{\rho }$, $b$, and $c$ being coupling constants.

Equations of motion for the mesonic fields are treated in the mean-field approximation with the solutions
\begin{eqnarray}
&& \rho^a_\mu = \delta_{a3}\delta_{\mu 0}\frac{g_{\rho N}}{2m_\rho^2}(n_p-n_n),
\quad
\om_\mu=\delta_{\mu 0}\frac{g_{\om N}}{m_\om^2}(n_p + n_n)\,,
\nonumber\\
&& m^2_\sigma \sigma + U'(\sigma)=g_{\sigma}\, (n_{{\rm S},p} + n_{{\rm S},n}),
\label{eqmotion_s}
\end{eqnarray}
where $n_{{\rm S},i}$, $i=n,p$, is the nucleon scalar density, which can be expressed through the nucleon density $n_i$ as
$$n_{{\rm S},i}=3n_i [x_i(x_i^2+1)^{1/2} -\log((x_i^2+1)^{1/2}+x_i)]/(2 x_i^3),$$
where  $x_i=p_{\rmF,i}/m_N^*$, and $p_{\rmF,i}=(3\pi^2 n_i)^{1/3}$ is the Fermi momentum. The effective nucleon mass $m_N^*=m_N(1-f)$ depends on the $\sigma$ field expressed through the dimensionless  variable $f=g_{\sigma}\sigma/m_N$.
Then the energy density following from the Lagrangian (\ref{Lag}) is given by
\begin{eqnarray}
&&E(n_p,n_n,f)=E_{\rm kin}(n_p,n_n,f) + E_V(n_p,n_n) + E_\sigma(f)\,,
\nonumber\\
&&E_{\rm kin}(n_p,n_n,f) = (\int\limits_{0}^{p_{{\rm F}n}} + \int\limits_{0}^{p_{{\rm F}p}}) \frac{p^2 dp}{\pi^2} \sqrt{p^2 + m_N^{*2}(f)}\,,
\nonumber\\
&&E_V(n_p,n_n)=\frac{C^2_\om}{2m_N^2}(n_p+n_n)^2 + \frac{C^2_\rho}{8m_N^2}(n_p-n_n)^2\,,
\nonumber\\
&&E_\sigma(f)=\frac{m_N^4f^2}{2C_\sigma^2}+U(f)\,,\quad  m_N^*(f)=m_N\,(1-f)\,,
\label{Eparts}
\end{eqnarray}
$C_j=g_j m_N/m_j$ with  $j=\sigma$, $\om$, $\rho$. In terms of the $f$ variable $U(f)=m_N^4 (b\,f^3/3+c\,f^4/4)$.
The pressure is given by:
\begin{eqnarray}
&& P(n_n, n_p, f) = P_{\rm kin}(n_n, n_p, f) + P_{V}(n_n, n_p) + P_\sigma(f),
\nonumber \\
&& P_{\rm kin}(n_n, n_p, f) = \frac{1}{3}(\int\limits_{0}^{p_{{\rm F}n}} + \int\limits_{0}^{p_{{\rm F}p}}) \frac{p^2 dp}{\pi^2} \frac{p^2}{\sqrt{p^2 + m_N^{*2}(f)}}\,, \nonumber \\
&& P_V(n_n, n_p) = E_V(n_p,n_n)\,,\quad P_\sigma(f) = - E_\sigma(f)\,.
\label{Pparts}
\end{eqnarray}
Five model parameters $C_\om$, $C_\sigma$, $C_\rho$, $b$ and $c$ are tuned to fit the values of  the saturation density $n_0$, the binding energy per nucleon
$\mathcal{E}_0=E(n_0/2,n_0/2,f_0)/n_0-m_N$, the compressibility modulus $K(n_0)$, the effective nucleon mass $m_N^*(n_0)=m_N(1-f_0)$, and the symmetry energy $E_{\rm sym}=\frac{n_0}{2}(\partial^2 E/\partial n_n^2)|_{n_p=n_n=n_0/2}$. The system of linear equations relating the saturation parameters and the parameters of the model can be found, e.g., in Ref.~\cite{Glendenning}. In our study we use as a reference the set of parameters: $n_0=0.16$~fm$^{-3}$ and $\mathcal{E}_0 = -16$~MeV and
\begin{eqnarray}
K(n_0) = 250 ~{\rm MeV},\, E_{\rm sym} = 30 ~{\rm MeV},\, f_0 = 0.2\, .
\label{param}
\end{eqnarray}
The parameters of our NLW model are $C^2_\sigma=196.343$, $C^2_\om=90.7682$,
$C^2_\rho=88.7261$, $b=8.94551\cdot 10^{-3}$, and
$c=7.70766\cdot 10^{-3}$, the nucleon mass is fixed as $m_N=938$\,MeV.

The composition of the NS matter  being in the $\beta$-equilibrium is governed by the relation between the lepton and nucleon chemical potentials $\mu_e=\mu_\mu=\mu_n-\mu_p$, where
$\mu_{n,p}=\partial E/\partial n_{n,p}$ and by the charge neutrality condition requiring that the charge of protons is compensated for by the charges of electrons and muons, i.e., $n_p=n_e+n_\mu$,  with $n_l= \theta(\mu_l^2-m_l^2)(\mu_l^2-m_l^2)^{3/2}/(3\pi^2)$ and $m_l$ standing for the mass of the lepton $l$.
Solution of this system of equations determines nucleon and lepton densities as functions of the total nucleon density $n=n_p+n_n$.

The total energy density and pressure, $E$ and $P$, for the NS matter are obtained after adding the kinetic contributions from leptons to the nucleon energy density and the pressure given by Eqs.~(\ref{Eparts}) and (\ref{Pparts}).
With $P = P(E)$ at hand, the NS masses and radii are calculated with the help of the Tolman-Oppenheimer-Volkoff (TOV) equation. In the crust region we match smoothly our EoS with the Baym-Pethick-Sutherland (BPS) EoS~\cite{BPS}. The limiting mass of a NS for the present choice of parameters (\ref{param}) is equal to $1.92~M_\odot$, which falls below the experimental limit.

Now we propose a simple method for controlling the dependence of the scalar field $f$ on the density that, as we demonstrate later, allows us to stiffen the EoS at densities higher than a certain chosen density $n_{*} > n_0$.
To be specific let us consider isospin symmetric matter (ISM).  The function $f(n)$ depends on the derivative of the scalar-field potential $U'(f)$, see Eq.~(\ref{eqmotion_s}). Hence by changing
\begin{eqnarray}U(f)\to \widetilde{U}(f)=U(f)+\Delta U(f)
\label{Utilde}
\end{eqnarray}
with an appropriately chosen $\Delta U(f)$ we may quench the $f(n)$ growth at densities $n\gsim n_{*}$. The requirement for the $\widetilde{U}(f)$ can be easily educed from the derivative of the scalar field over the nucleon density following from Eq.~(\ref{eqmotion_s})
\begin{eqnarray}
\frac{\rmd f}{\rmd n} = \frac{2 ({\partial n_{\rm S}}/{\partial n})}
{m_N^3C_\sigma^{-2} + \widetilde{U}''(f)/m_N - 2 ({\partial n_{\rm S}}/{\partial f})},
\label{dsdpf}
\end{eqnarray}
where $\widetilde{U}''(f)$ stands for the second derivative of the potential $\widetilde{U}$ with respect to $f$.
The partial derivatives of the scalar density are equal to
\begin{eqnarray}
\frac{\partial n_{\rm S} }{\partial n} = \frac{m_N^*}{2\sqrt{p_\rmF^2 + m_N^{*2}}},
\quad
-\frac{\partial n_S}{\partial f} =   \int\limits_{0}^{p_\rmF} \frac{m_N p^4 \rmd p/\pi^2} {(p^2 + m_N^{*2})^{3/2}}\,.
\end{eqnarray}
From Eq.~(\ref{dsdpf}) we see that although $\rmd f/\rmd n$ is always positive [provided $\widetilde{U}''(f)>0$] it can be made as small as required if one takes $\widetilde{U}''(f)$  sufficiently large for $f$ in the vicinity of a  value $f_{\rm core}$ determined by the choice of the density $n_{*}$.
According to Eq.~(\ref{dsdpf}) the requirement of a slow growth of the function $f(n)$, $n\rmd f / \rmd n \ll 1$  translates into the condition
\begin{eqnarray}
\Delta{U}''(f)\gg {m_N m_N^*\,n}/{\sqrt{p_\rmF^2+m_N^{*2}}},
\label{ineq}
\end{eqnarray}
which is insensitive to the particular choice of the original potential ${U}(f)$.

\begin{figure}
\includegraphics[width=8.6cm]{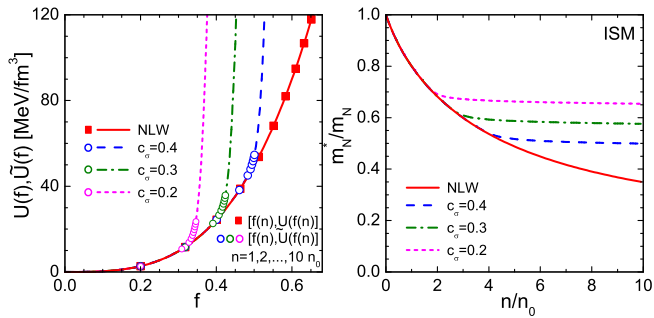}
\caption{Left panel: the scalar field potential $U(f)$  in the NLW model and
$\widetilde{U}(f)$ in the NLWcut model  for various values $c_\sigma$,  as  functions of the scalar field parameter $f$. The symbols indicate the values of $[f(n); {U}(f(n)),\widetilde{U}(f(n))]$ for the density $n$ of the ISM varying from $n_0$ to $10n_0$ with the $1n_0$ step.
Right panel: the nucleon effective mass in the ISM as a function of the nucleon density for the same models.  }
\label{fig:meff-U}
\end{figure}

Now let us give two examples of the appropriate dependence of $\Delta{U}(f)$. We will refer to such an extension as the ``$\sigma$-cut scheme'' and denote the model as the NLWcut model.
The potential
\begin{eqnarray}
&&\Delta{U}(f)=\alpha\ln[1+\exp(\beta(f-f_{\rm s.core}))]
\label{Umod}
\end{eqnarray}
allows $f$ to grow above the $f_{\rm s.core}$. This could be interpreted as an inclusion of the excluded volume effects with a soft core in the scalar baryon density.
The maximum of $\Delta U''(f)$ in Eq.~(\ref{Umod}) is realized for $f=f_{\rm s.core}$ and is equal to $\alpha\beta^2/4$. Hence condition (\ref{ineq}) means
$\alpha\beta^2/(4 m_N^4) \gg 1.5\times 10^{-3}(n/n_0)$.
For  numerical illustrations  we  use the potential (\ref{Umod}) with the parameters $\alpha= m_\pi^4= 4.822\times 10^{-4}\,m_N^4$ and $\beta=120$ and vary the value $f_{\rm s.core}$.
Thus, we have $\alpha\beta^2/(4 m_N^4) \simeq 1.7$, and inequality (\ref{ineq}) is fulfilled.
If we choose a singular potential diverging at $f=f_{\rm h.core}$,
\begin{eqnarray}
\Delta{U}(f)=\alpha\big[{\delta f}/{(f_{\rm h.core}-f)}\big]^{2\beta}\,,
\label{U-hardcore}
\end{eqnarray}
the function $f(n)$ will never exceed the value $f_{\rm h.core}$. This could be interpreted as an inclusion of the excluded volume effects with a hard core in the scalar baryon density.
For densities $n<10n_0$ the same dependence $f(n)$ as for the potential (\ref{Umod}) can be recovered with the potential in the form (\ref{U-hardcore}) with parameters $\alpha=0.809\,m_\pi^4$, $\beta=11.138$, and $f_{\rm h.core}= f_{\rm s.core}+\delta f$ with $\delta f=0.190$.

Applying the method to models with different values of $f_0$
it is convenient to present
\begin{eqnarray}
f_{\rm s.core}=f_0+c_\sigma(1-f_0)\,.
\label{fcore}
\end{eqnarray}
For illustration we use the values $c_\sigma = 0.2, 0.3, 0.4$, which correspond to
$f_{\rm s.core}\approx 0.36$, $0.44$, and $0.52$ for $f_0 =0.2$.

The  potentials $U$ and  $\widetilde{U}$  given by Eq.~(\ref{Umod}) are shown on the left panel in Fig.~\ref{fig:meff-U} as a function of $f$. Symbols mark the values of $[f(n);U(f(n))]$ and  $[f(n);\widetilde{U}(f(n))]$ for various densities $n$. We see that indeed the symbols pile up in the region where the potential $\widetilde{U}$  has a maximal rise.
On the right panel in Fig.~\ref{fig:meff-U} we show the effective nucleon mass $m_N^*(n)=m_N(1-f(n))$ as a function of the nucleon density, where $f(n)$ follows as a solution of Eq.~(\ref{eqmotion_s}). We see that the function $m_N^*(n)$ flattens off for densities $n>n_{*}\simeq 1.9\,n_0$, $3.0\,n_0$, and $4.0\,n_0$ corresponding to $c_\sigma=0.2$, $0.3$, and $0.4$, respectively. The closer the $f_{\rm s.core}$ value is to $f_0$, i.e. the smaller $c_\sigma$ is, the smaller the density is, at which the NLWcut model starts to deviate from the original NLW model. The NLW model is recovered from the  NLWcut model for $c_\sigma\gsim 1$, when $\widetilde{U}(f)$ almost coincides with $U(f)$ for  $0\le f\lsim 1$ for all densities.
We stress that all saturation properties of the NLWcut models remain the same as for the original NLW model, since by construction $\Delta U (n_0)$ is chosen to be a tiny value.

\begin{figure}
\includegraphics[width=8.6cm]{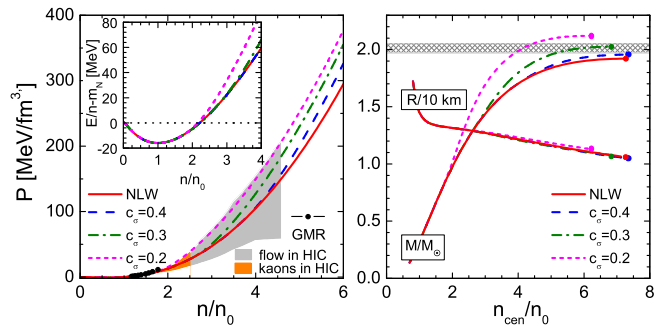}
\caption{Left panel: Total pressure $P$ for the NLW and NLWcut models as a function of the nucleon density in the ISM for various values $c_{\sigma}$; the insertion demonstrates the binding energy per nucleon. Two shadowed areas show the constraints on the pressure from the particle flow and the kaon production data in HIC extracted in~\cite{HICflow,Lynch09}. Bold dots show the extrapolation of the pressure consistent with GMR~\cite{Lynch09}.
Right panel: The NS mass and radius as functions of the central density for our models. The hatched band denotes the uncertainty in the value of the maximum measured NS mass of $(2.01 \pm 0.04)~M_\odot$.}
\label{fig:parts_sigma}
\end{figure}

The influence of the quenching of the nucleon-mass decrease on an EoS is illustrated in the left panel in Fig.~\ref{fig:parts_sigma}, where we show the total pressure, Eq.~(\ref{Pparts}), and the binding energy per nucleon as functions of the nucleon density in the ISM. We observe that the replacement (\ref{Umod}) leads to a sizable increase of the pressure for $n>n_{*}$ because of a reduction of the negative contribution to the pressure coming from the explicitly $\sigma$-field-dependent term, $P_\sigma$, which overcomes a slight decrease of the effective kinetic term $P_{\rm kin}$ owing to a relative increase of the nucleon mass $m_N^*$. Constraints on the pressure extracted from the particle flow and the kaon production in heavy-ion collisions (HICs)~\cite{HICflow,Lynch09} are shown by the shaded areas and from analysis of the data on the giant monopole resonance (GMR)~\cite{Lynch09}, by the bold dots. We see that the EoSs in the NLWcut model with the input parameters (\ref{param}) satisfy these constraints for $c_\sigma > 0.2$, exceeding the upper limit of the HIC constraints only slightly for  $c_\sigma=0.2$ and in a narrow interval of densities, $2\,n_0<n<2.6\,n_0$.

Different from the pressure, the energy-density and corresponding binding energy increase more weakly after the inclusion of the new term  (\ref{Umod}) to the potential. This occurs because an increase in the effective kinetic energy $E_{\rm kin}$ density is almost fully compensated for by a decrease in the explicitly $\sigma$-field dependent contribution $E_\sigma$.
As a consequence, the net effect of the replacements (\ref{Utilde}) and (\ref{Umod}) is a stiffening of the EoS $P(E)$ for densities $n>n_{*}$. This result holds also for pure neutron matter and NS matter since the symmetry energy is not influenced by the replacement (\ref{Utilde}).

The stiffening of the EoS is reflected in an increase in the NS masses as functions of the central density, which are shown in the right panel in Fig.~\ref{fig:parts_sigma} for our models together with the experimental constraint $M = (2.01 \pm 0.04)\,M_\odot$. The proposed $\sigma$-cut scheme allows us to shift the maximal NS mass from $1.92\,M_\odot$ for the original NLW model to $1.96\,M_\odot$, $2.03\,M_\odot$, and $2.12\,M_\odot$ for $c_\sigma=0.4$, $0.3$, and $0.2$, respectively. Thus, using our modification one is able to fit the experimental constraint on the maximum NS mass, provided the value $f_{\rm s.core}$ is chosen low enough (corresponding to $c_\sigma = 0.3$ and $0.2$ in our example). Also, the NS radius $R$ is shown in this panel. As one may see, $R$ changes only slightly with $c_\sigma$ increasing  with a decrease of $c_\sigma$.
For example, within the NLWcut models with $c_\sigma\geq 0.4$ a NS with the mass $M= 1.4 M_{\odot}$ has radius $R=12.7$\,km, whereas $R=12.9$\,km for $c_\sigma=0.2$.

\begin{figure}
\begin{center}
\includegraphics[width=8.6cm]{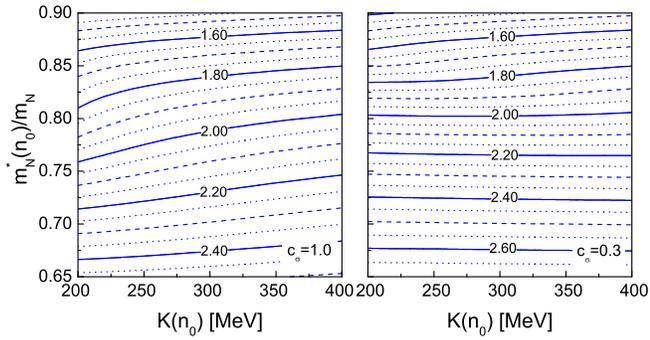}
\end{center}
\caption{Contour plot of the maximum NS mass as a function of $m_N^*(n_0)$ and $K(n_0)$ for the   NLWcut models with $c_{\sigma}=1$, simulating the case of the  original NLW model (left panel) and $c_\sigma=0.3$ (right panel). Contour lines are shown with the mass step $0.05\,M_\odot$.}
\label{fig:contour_M}
\end{figure}

It is instructive to study the dependence of the maximum NS mass on the input values
$m_N^*(n_0)$ and $K(n_0)$ for different cut parameters $c_\sigma$. In Fig.~\ref{fig:contour_M} we show the contour plot of the maximum NS mass for the original NLW model and the NLWcut model with
$c_\sigma=0.3$. We see that the application of the $\sigma$-cut scheme removes completely the
dependence of $M_{\rm max}$ on the compressibility modulus $K(n_0)$ for $m_N^*(n_0)/m_N<0.8$, the
contour lines on the right panel in Fig.~\ref{fig:contour_M} are almost parallel to the $K(n_0)$ axis. The gain in $M_{\rm max}$ proves to be larger, the smaller the ratio $m_N^*(n_0)/m_N$ is, reaching, e.g., at $m_N^*(n_0)/m_N\simeq 0.65$, the value of $0.2\, M_\odot$ for $c_\sigma=0.3$ and $0.35\,M_\odot$ for $c_\sigma=0.2$.

\begin{figure}
\includegraphics[width=8.6cm]{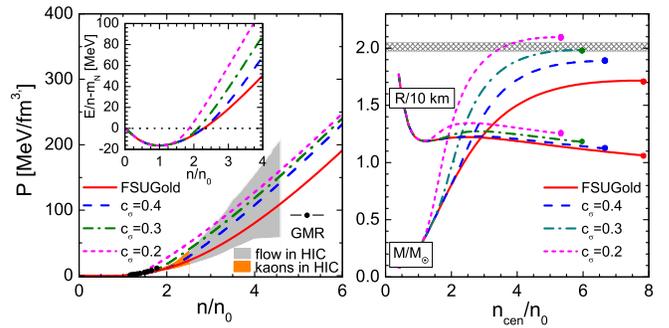}
\caption{
Same as Fig.~\ref{fig:parts_sigma} but for the FSUGold model.
}
\label{FSUmr}
\end{figure}

In this work we leave aside the question about hyperonization of the NS matter at high densities, which would soften the EoS and decrease the maximum available NS mass. A typical decrease of the maximum  NS mass owing to hyperonization is rather large, typically $\sim 0.3\mbox{--}0.4 M_{\odot}$, so that the maximum calculated NS mass may become below  the maximum measured NS mass (the so-called ``hyperon puzzle''~\cite{Hypprob}).  If included within the NLW model that we exploit here, the hyperons would appear at the density $2.73\,n_0$, a value that changes weakly, if the $\sigma$-cut scheme is applied, decreasing with an decrease of $c_\sigma$ up to $2.56\,n_0$ for $c_\sigma=0.2$. A possible mechanism to overcome the problem within the $\sigma$-cut scheme presented in this work is the use of a smaller values of $m_N^*(n_0)$ and/or $c_\sigma$, see Fig.~\ref{fig:contour_M}. However, in this case the problem is solved at the price of a possible violation of the HIC constraints on the pressure in a certain density interval. Other mechanisms to solve the puzzle might be an alternative choice of the coupling constants, see~\cite{Weissenborn:2011ut}, field- (or density)-dependent parameters of the model~\cite{Typel,MKVshort}, or something else.

As a further illustration we apply the $\sigma$-cut scheme to the FSUGold model~\cite{Toddrutel}, which proved to be fine tuned to describe the properties of finite nuclei. This model includes the vector meson self-interactions, which are described by the interaction Lagrangian:
$
\mathcal{L}_{\rm int} = \zeta(g_{\om}^2 \om_\mu \om^\mu)^2/4! + \Lambda_{v} g_{\om}^2 g_\rho^2 \om_\mu \om^\mu \vec{\rho}_\nu \vec{\rho}^\nu,
$
with additional parameters $\zeta$ and $\Lambda_{v}$ determined as in~~\cite{Toddrutel}.  We modify the model by adding the scalar-field potential (\ref{Umod}) with the same values of $\alpha$ and $\beta$  and consider $c_\sigma = 0.2, 0.3, 0.4$.

The left panel in Fig.~\ref{FSUmr} shows the binding energy per nucleon and the pressure in the ISM for the FSUGold model and its $\sigma$-cut extension.
The energy and pressure for the $\sigma$-cut model begin to differ significantly from the FSUGold model at densities $n \simeq 1.8\,n_0$ for $c_\sigma =0.3$ and $n\simeq 1.4\, n_0$ for $c_\sigma =0.2$. The saturation properties of the FSUGold model are almost not affected for $c_\sigma \gsim 0.2$. However, for $c_\sigma=0.2$ the pressure exceeds the HIC and GMR constraints at densities $n<3 n_0$. For $c_\sigma=0.3$ the constraints are satisfied except for a narrow density interval $1.7 n_0<n<2.7 n_0$.
In the right panel in Fig.~\ref{FSUmr} we show the NS mass and radius  as  functions of the central density for the original FSUGold model and the FSUGold model with the scalar-field potential modified as in Eqs.~(\ref{Utilde}) and (\ref{Umod}) for various values of $c_\sigma$. The application of the $\sigma$-cut scheme allows us to increase the maximum NS mass from the original $1.72\,M_\odot$ up to $1.89\,M_\odot$, $1.98\,M_\odot$ and $2.09~M_\odot$ for $c_\sigma=0.4$, $0.3$, and $0.2$, respectively. So, the model starts to satisfy the heavy NS mass constraint  for $c_\sigma\lsim 0.3$. Thus the re-tuning of the FSUGold model performed in~\cite{FSUGold2} would be unnecessary, if one exploited the $\sigma$-cut scheme. We note that the increase of the maximum NS mass is larger in the FSUGold model than in the NLW model with parameters (\ref{param}) this can be related to the choice of the smaller value of $m_N^*(n_0)=0.61\, m_N$ in the former model.
The NS radius for the FSUGold model is more sensitive to the $\sigma$ cut than for the NLW model, increasing for $M=1.4 M_{\odot}$ from 12.0 to $12.3$, $12.7$, and $13.3$\,km for $c_\sigma=0.4$, 0.3, and 0.2, respectively.

In conclusion, we proposed a simple method of making the EoS obtained in a relativistic mean-field model stiffer at densities larger than some chosen value $n_{*}$. The strategy is to quench the growth of the scalar field at the value $f_{\rm s.core}$. This can be achieved by adding to the energy-density  a functional of the scalar field, which is vanishingly small for $f \sim f_0 < f_{\rm s.core}$ and is rapidly increased  for $f\sim f_{\rm s.core}$, so that its second derivative becomes sufficiently large. Then, the self-consistent solution of the corresponding equation of motion does not let the function $f(n)$ to grow significantly above $f_{\rm s.core}$ within a broad interval of densities, for $n>n_{*}$. Herewith the EoS remains unaltered for densities $n\sim n_{0}$ but stiffens significantly for $n>n_{*}$, which results in an increase of the limiting value of the neutron star mass. We demonstrated the work of this method at hand of the standard non-linear Walecka model. Then we applied the method to the FSUGold EoS and showed that it is possible to increase the maximum neutron star mass bringing it in agreement with modern astrophysical constraints, without changing the EoS for densities below $n\sim n_0$.

The value $f_{\rm s.core}$ might be associated  with a manifestation of the nuclear core in the medium and for $f$ in the vicinity of $f_{\rm s.core}$ one may expect  occurrence of the deconfinement phase transition.


This work was supported by the Ministry of Education and Science
of the Russian Federation (Basic part), by the Slovak Grants No.~APVV-0050-11 and No.~VEGA-1/0469/15, and by ``NewCompStar'', COST
Action MP1304. E.E.K and K.A.M. acknowledge the hospitality of the Joint Institute of Nuclear Research in Dubna (Russia), where part of this work was done. E.E.K acknowledges the support
by the Plenipotentiary of the Slovak Government to JINR Dubna.
Computing was partially performed in the High Performance Computing Center of the Matej Bel University using the HPC infrastructure acquired in Project ITMS 26230120002 and 26210120002 (Slovak infrastructure for high-performance computing) supported by the Research \& Development Operational Programme funded by the ERDF.

\end{document}